\documentclass[prl,reprint,english,twocolumn,showpacs,superscriptaddress]{revtex4-1}
\usepackage{bm,mathrsfs}
\usepackage[dvipdfmx]{graphicx}
\usepackage{epsfig}
\usepackage{psfrag}
\usepackage{amsmath,bbm}
\usepackage{amsfonts,amssymb}
\usepackage[utf8]{inputenc}
\usepackage[T1]{fontenc}
\usepackage{times}
\usepackage{dsfont}
\usepackage{enumitem}  
\usepackage{comment}
\usepackage[dvipdfmx,colorlinks=true,linkcolor=blue,urlcolor=blue,citecolor=blue]{hyperref}
\usepackage{braket,physics}
\usepackage{verbatim,color,ulem}
\newcommand{\beq}{\begin{equation}}
\newcommand{\eeq}{\end{equation}}
\newcommand{\bseq}{\begin{subequations}}
\newcommand{\eseq}{\end{subequations}}
\newcommand{\bal}{\begin{aligned}[b]}
\newcommand{\eal}{\end{aligned}}
\newcommand{\beqa}{\begin{eqnarray}}
\newcommand{\eeqa}{\end{eqnarray}}

\begin{document}

\title{Interaction-induced dissipative quantum phase transition \\in a head-to-tail atomic Josephson junction}

\author{Koichiro Furutani}
\email{furutani.koichiro.s7@f.mail.nagoya-u.ac.jp}
\affiliation{Department of Applied Physics, Nagoya University, Nagoya 464-8603, Japan}
\affiliation{Institute for Advanced Research, Nagoya University, Nagoya 464-8601, Japan}
\author{Luca Salasnich}
\affiliation{Dipartimento di Fisica e Astronomia 'Galileo Galilei'  
and QTech Center, Universit\`a di Padova, via Marzolo 8, 35131 Padova, Italy}
\affiliation{Istituto Nazionale di Fisica Nucleare, Sezione di Padova, 
via Marzolo 8, 35131 Padova, Italy}
\affiliation{Istituto Nazionale di Ottica del Consiglio Nazionale delle Ricerche, 
via Carrara 2, 50019 Sesto Fiorentino, Italy}

\date{\today}

\begin{abstract}
We propose a dissipative phase transition in a head-to-tail Bose Josephson junction. 
The quantum phase transition has the same origin as the one in a resistively shunted Josephson junction, but the intrinsic momentum coupling between the Josephson mode and the bath modes enables us to observe the dissipative phase transition without any synthetic dissipation.  
We show that the interatomic interaction strength plays the role of the damping parameter. 
Consequently, in contrast to a resistively shunted Josephson circuit, the Bose Josephson junction can exhibit an insulating phase in a wider parameter region by increasing the repulsive interaction strength, which is robust against nonperturbative effects. 
We argue that tight transverse confinement of the quasi-one-dimensional atomic gas allows us to reach the insulating phase. 
\end{abstract}

\maketitle

Understanding of the dissipative nature of quantum systems is increasing in importance for the manipulation of quantum devices. 
Particularly, a dissipative phase transition in a resistively shunted Josephson junction (RSJJ) proposed by Schmid and Bulgadaev 40 years ago has again attracted intense interest recently \cite{schmid1983,bulgadaev1984}. 
A RSJJ can be described by the Caldeira-Leggett (CL) model with a Josephson cosine potential \cite{caldeira1983}. 
The CL model, composed of a single particle subject to an external potential coupled with phonons of a thermal bath, is one of the most fundamental setups to analyze quantum dissipation in quantum many-body systems. 
In a RSJJ, the resistor causes thermal noise and friction, and the damping parameter is determined by the resistance \cite{koch1980,kf2021}. 
Within the perturbative renormalization group (RG) analysis, the capacitive contribution was considered irrelevant and a dissipative phase transition exactly at the quantum resistance was predicted irrespective of the Josephson energy and the charging energy, which had, however, never been experimentally verified for a long time \cite{murani2020}. 
In contrast to the conventional understanding established by Schmid and Bulgadaev, in recent studies, the capacitive contribution turned out to be relevant in the nonperturbative regime and it crucially modifies the phase diagram suppressing the insulating phase dramatically \cite{masuki2022,yokota2023}. 
It implies that the effects of dissipation in the CL model require nonperturbative treatments to capture the quantum phase transition, which is theoretically tough and complicated to analyze. 
Even though these highly developed nonperturbative approaches are widely useful in general, they can also result in discrepancies among different approaches \cite{troyer2005,werner2007,murani2020,masuki2022,dupuis2023,snyman2023,masuki2023rep,joyez2023,glazman2024}, 
which causes considerable difficulty in comprehensive interpretations. 
Then, a natural concern is whether we need to always examine the nonperturbative contributions to determine the ground state of these dissipative systems. 

In this Letter, we answer this issue by proposing a physical system exhibiting the Schmid-Bulgadaev dissipative phase transition robust against nonperturbative effects: an atomic Bose Josephson junction (BJJ) in a head-to-tail configuration. 
The atomic Josephson junction is described by a CL-type model even without any extrinsic coupling with a reservoir and the relative phase obeys a generalized quantum Langevin equation \cite{minguzzi2018,binanti2021,furutani2023}. 
We call it an intrinsically-momentum-coupled CL (ICL) model. 
The ICL model hosts qualitatively distinct properties from the standard CL model. 
Within the ICL model, we point out that the dissipative phase diagram recovers the Schmid-Bulgadaev picture, which was originally derived by a perturbative approach, even {\it beyond} the perturbative regime. 
It is owing that the ICL model with the Josephson coupling is equivalent to the boundary sine-Gordon model at any parameter region due to the intrinsic momentum coupling between the Josephson mode and the bath modes in contrast to the RSJJ. 
Based on the phase diagram, we predict a dissipative phase transition from the superfluid phase to the insulating phase, which is broadened rather than the one in the RSJJ. 
Any external dissipator such as a resistor in an electric circuit is no longer necessary in this dissipative phase transition, but the ground state is controlled by the inter-atomic interaction strength. 
In our head-to-tail BJJ, reaching the quantum phase transition turns out to require a large gas parameter. 
We argue that tight constrictions in the transverse directions to realize a quasi-one-dimensional BJJ have a possibility to observe the dissipative phase transition due to the renormalized interaction.

\begin{figure}[tb]
\centering
\includegraphics[keepaspectratio,scale=0.13]{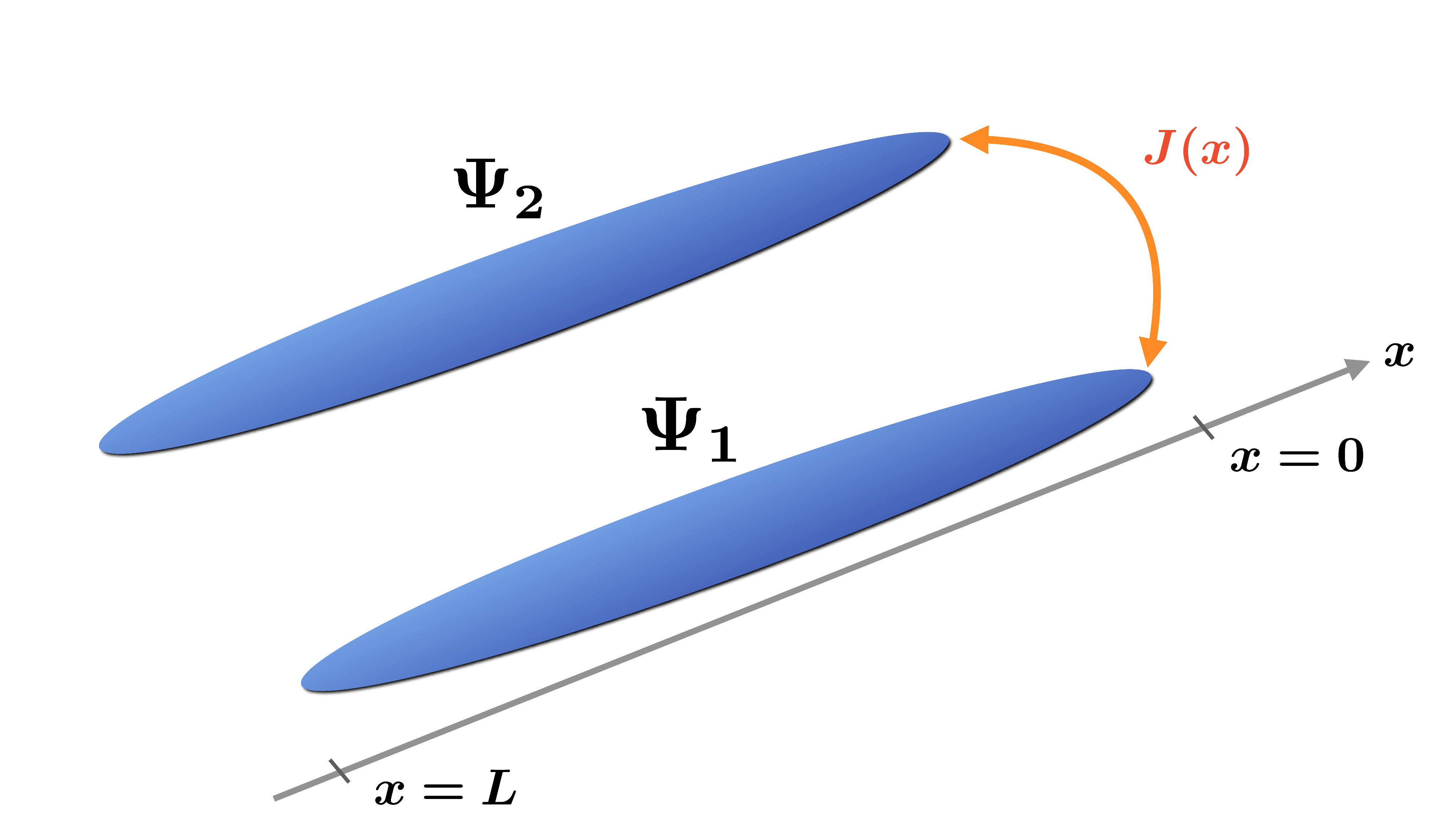}
\caption{Bose Josephson junction in the head-to-tail configuration with $J(x)=J_{0}L\delta(x)$. 
Two quasi-one-dimensional Bose gases of system size $L$ described by the complex Bose fields $\Psi_{1,2}(x,t)$ are coupled through the Josephson coupling $J(x)$ at $x=0$. }
\label{FigBJJ}
\end{figure}

We start from a quasi-one-dimensional atomic BJJ in a head-to-tail configuration with the system size $L$ described by 
\beq
\bal
\mathscr{L}_{\rm BJJ}&=\sum_{a=1,2}\left[i\hbar\Psi_{a}^{*}\partial_{t}\Psi_{a}-\dfrac{\hbar^{2}}{2m}\abs{\partial_{x}\Psi_{a}}^{2}-\frac{g}{2}\abs{\Psi_{a}}^{4}\right] \\
&+\dfrac{J(x)}{2}\left[\Psi_{1}^{*}\Psi_{2}+\Psi_{2}^{*}\Psi_{1}\right],
\eal
\label{LBJJ}
\eeq
where $\Psi_{a=1,2}(x,t)=\sqrt{n_{a}(x,t)}e^{i\phi_{a}(x,t)}$ is the complex Bose field of each tube, $m$ is the atomic mass, $g$ is the $s$-wave interaction strength, and $J(x)=J_{0}L\delta(x)$ is the Josephson coupling. 
The schematic picture of the system is depicted in Fig.~\ref{FigBJJ}. 
To analyze the relative dynamics, we introduce a relative phase $\phi=\phi_{1}-\phi_{2}$ and a population imbalance $\zeta=(n_{1}-n_{2})/(n_{1}+n_{2})$, and perform mode expansions as \cite{binanti2021}
\beq
\phi(x,t)=\dfrac{1}{\sqrt{L}}\sum_{n=0}^{N}\Phi_{n}(x)q_{n}(t), 
\eeq
with $\{\Phi_{n}(x)\}=\{\cos{(k_{n}x)}\}/\sqrt{L}$ being the orthonormal basis and $k_{n}=n\pi/L$ with $n=0,1,2,\cdots N$. 
The mode number cutoff can be chosen to be the total number of atoms in atomic systems \cite{binanti2021}. 
One can then identify the Josephson mode $\phi_{0}(t)=\sum_{n=0}^{N}q_{n}(t)/L$ and the bath mode $Q_{n}(t)=q_{n}(t)$ for $n\ge1$. 
Under this head-to-tail configuration with a small population imbalance $\abs{\zeta}\ll1$ and $\partial_{x}^{2}\zeta\simeq 0$, a canonical transformation maps the BJJ described by Eq.~\eqref{LBJJ} to \cite{binanti2021}
\beq
\bal
H_{\rm ICL}&=\frac{P_{0}^{2}}{2M}-J_{0}L\bar{n}\cos{\phi_{0}} \\
&+\sum_{n=1}^{N}\left[\frac{(P_{n}+P_{0})^{2}}{2M}+\frac{M\omega_{n}^{2}}{2}Q_{n}^{2}\right], 
\eal
\label{Hphi}
\eeq
where $(Q_{n},P_{n})$ ($n\ge1$) is a set of coordinate and canonical momentum of the bath modes, and $(L\phi_{0},P_{0})$ is the coordinate and canonical momentum of the Josephson mode with the effective mass $M=\hbar^{2}/2gL$, $\bar{n}=(\abs{\Psi_{1}}^{2}+\abs{\Psi_{2}}^{2})/2=N/L$, and $\omega_{n}=c k_{n}$. 
The sound velocity is given by $c=\sqrt{g\bar{n}/m}$. 
The Hamiltonian \eqref{Hphi} describes a coherent Josephson mode coupled with incoherent Bogoliubov phonons, and they are coupled intrinsically through the canonical momentum. 
This intrinsic coupling between the Josephson mode and bath modes leads to dissipation and the Josephson mode obeys a generalized Langevin equation \cite{minguzzi2018,binanti2021}. 
We call Eq.~\eqref{Hphi} the ICL model. 
Within the assumptions of a small population imbalance $\abs{\zeta}\ll1$ and $\partial_{x}^{2}\zeta\simeq 0$, the population imbalance is proportional to the time derivative of the relative phase \cite{binanti2021}. 
The damped dynamics of the relative phase can therefore yield the assumed small population imbalance and negligible higher-order spatial derivatives consistently. 
Moreover, the magnitude of fluctuations in the population imbalance is fairly suppressed compared to phase fluctuations due to the intrinsic momentum coupling, which also supports the assumptions. 
Second quantization with $Q_{n}\to\Hat{Q}_{n}=i\sqrt{\hbar/2M\omega_{n}}\, (\Hat{b}_{n}^{\dagger}-\Hat{b}_{n})$ and $P_{n}\to\Hat{P}_{n}=-i\sqrt{\hbar M\omega_{n}/2}\, (\Hat{b}_{n}^{\dagger}+\Hat{b}_{n})$
with $\Hat{b}_{n}^{\dagger}$ and $\Hat{b}_{n}$ being the creation and annihilation operators of the $n$th bath modes, respectively, satisfying the commutation relation $[\Hat{b}_{n},\Hat{b}_{m}^{\dagger}]=\delta_{nm}$, and $\Hat{N}_{0}\equiv -i\partial_{\phi_{0}}=L\Hat{P}_{0}/\hbar$ results in $H_{\rm ICL}\to\Hat{H}_{\rm ICL}=\Hat{H}_{\rm J}+\Hat{H}_{\rm B}+\Hat{H}_{\rm JB}$ with
\bseq
\beq
\Hat{H}_{\rm J}=E_{C}\Hat{N}_{0}^{2}-E_{J}\cos{\phi_{0}},
\label{HhatJ}
\eeq
\beq
\Hat{H}_{\rm B}=\sum_{n=1}^{N}\hbar\omega_{n}\Hat{b}_{n}^{\dagger}\Hat{b}_{n},
\eeq
\beq
\Hat{H}_{\rm JB}=-\Hat{N}_{0}\sum_{n=1}^{N}\hbar\kappa_{n}\left(\Hat{b}_{n}^{\dagger}+\Hat{b}_{n}\right) ,
\eeq
\label{Hhat}
\eseq
with
\beq
E_{C}=\frac{(1+N)\hbar^{2}}{2ML^{2}}, \quad 
E_{J}=J_{0}L\bar{n}, \quad
\hbar \kappa_{n}=\sqrt{\frac{E_{C}\hbar\omega_{n}}{1+N}}. 
\label{ECEJkappa}
\eeq
The system described by $\Hat{H}_{\rm ICL}$ with Eqs.~\eqref{Hhat} is equivalent to a RSJJ \cite{masuki2022}. 
Note that, however, our BJJ model \eqref{Hhat} does not have any off-diagonal terms such as $\Hat{b}_{n}^{\dagger}\Hat{b}_{m}^{\dagger}$ in contrast to the RSJJ (see Eq.~(1) in Ref.~\cite{masuki2022}). 
Consequently, we do not need any diagonalization and the coupling has a simpler form. 
Instead, the charging energy $E_{C}$ involves the total number of atoms in each tube $N$, which determines the cutoff frequency $W=\omega_{N}=c \pi N/L$. 
The form of the coupling with the bath in Eq.~\eqref{ECEJkappa} in the BJJ yields a qualitative difference in the phase diagram from the RSJJ \cite{masuki2022}.

\begin{figure}[tb]
\centering
\includegraphics[keepaspectratio,scale=0.13]{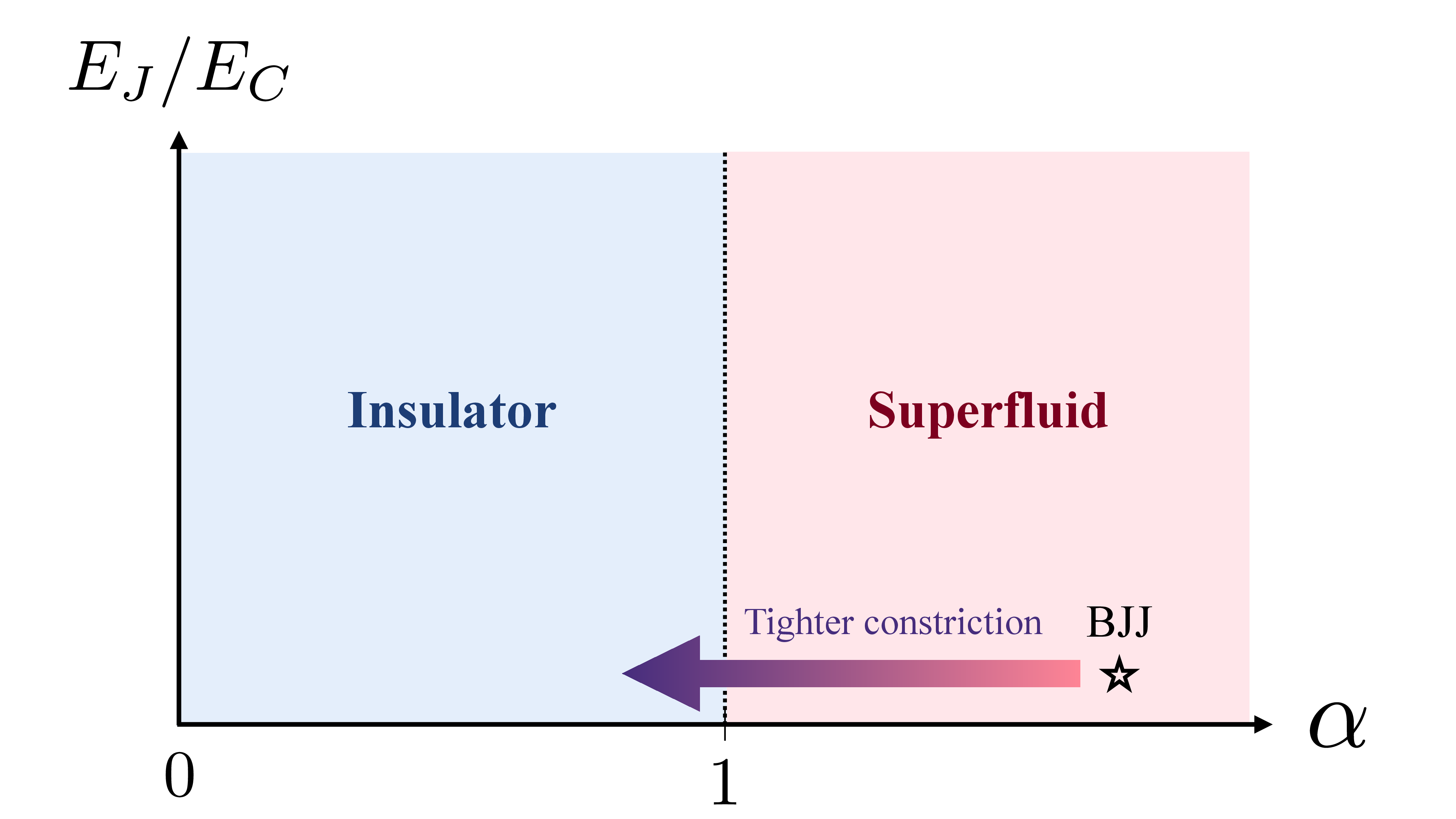}
\caption{Phase diagram of the ICL model with the Josephson coupling with respect to $\alpha=2\pi/\sqrt{\Tilde{g}}$ where $\Tilde{g}=mg/\bar{n}\hbar^{2}$ is the gas parameter. 
The vertical axis represents the ratio between the Josephson energy and the charging energy $E_{J}/E_{C}$. 
The region $\alpha>1$ corresponds to the superfluid state with the localized superfluid phase. 
The region $\alpha<1$ corresponds to the insulating state with the delocalized phase. }
\label{Figphase}
\end{figure}

To see that practically, we consider a canonical transformation of Eq.~\eqref{Hphi}, or alternatively, a unitary transformation of Eq.~\eqref{Hhat} as $\Hat{H}=\Hat{U}^{\dagger}\Hat{H}_{\rm ICL}\Hat{U}$ with $\Hat{U}\equiv \mathrm{exp}[-i\Hat{N}_{0}\Hat{\Xi}]$ and
$\Hat{\Xi}\equiv i\sum_{n=1}^{N}\kappa_{n}(\Hat{b}_{n}^{\dagger}-\Hat{b}_{n})/\omega_{n}$
\cite{masuki2022,ashida2021}. 
As a result, we can obtain the transformed Hamiltonian as \cite{masuki2022}
\beq
\Hat{H}=-E_{J}\cos{\left[\phi_{0}+\frac{1}{\sqrt{\alpha}}\Hat{\varphi}(0)\right]}
+\frac{\hbar c}{4\pi}\int^{L}_{0}\dd{x}\left[\left(\partial_{x}\Hat{\varphi}\right)^{2}+\Hat{\pi}^{2}\right], 
\label{bsg}
\eeq
with
\bseq
\beq
\Hat{\varphi}(x)\equiv \sum_{n=1}^{N}\sqrt{\frac{2\pi}{k_{n}L}}\,i\left(\Hat{b}_{n}^{\dagger}-\Hat{b}_{n}\right)\cos{(k_{n}x)},
\eeq
\beq
\Hat{\pi}(x)\equiv \sum_{n=1}^{N}\sqrt{\frac{2\pi k_{n}}{L}}\,\left(\Hat{b}_{n}^{\dagger}+\Hat{b}_{n}\right)\sin{(k_{n}x)}.
\eeq
\eseq
We introduced a parameter
\beq
\alpha
=\frac{2\pi}{\sqrt{\Tilde{g}}}. 
\label{alpha}
\eeq
where $\Tilde{g}=mg/\bar{n}\hbar^{2}$ is the gas parameter. 
Equation \eqref{bsg} is the boundary sine-Gordon model and the ground state is known to be classified as a superconducting state for $\alpha>1$ or an insulating state for $\alpha<1$ even beyond the perturbative regime \cite{masuki2022,schmid1983,bulgadaev1984,muramatsu1985,zwerger1985,fisher1992,nagaosa1993,saleur1995}. 
The phase diagram at zero temperature is summarized in Fig.~\ref{Figphase}. 
Indeed, the RG flow of the Josephson energy in the boundary sine-Gordon model reads \cite{altlandsimons,nagaosaQFT}
\beq
\partial_{\ell}\epsilon_{J}(\ell)=\left(1-\dfrac{1}{\alpha}\right)\epsilon_{J}(\ell) ,
\label{EJRG}
\eeq
at the lowest order with $\ell$ being the dimensionless RG scale and $\epsilon_{J}=E_{J}/\hbar W$. 
The RG cutoff scale can be determined by the frequency cutoff as $\ell_{\rm max}=\ln{(W/\omega_{1})}$. 
It is related to the total number of atoms $N$ in a cold-atom setup \cite{binanti2021}. 
\begin{figure}[t]
\includegraphics[keepaspectratio,scale=0.32]{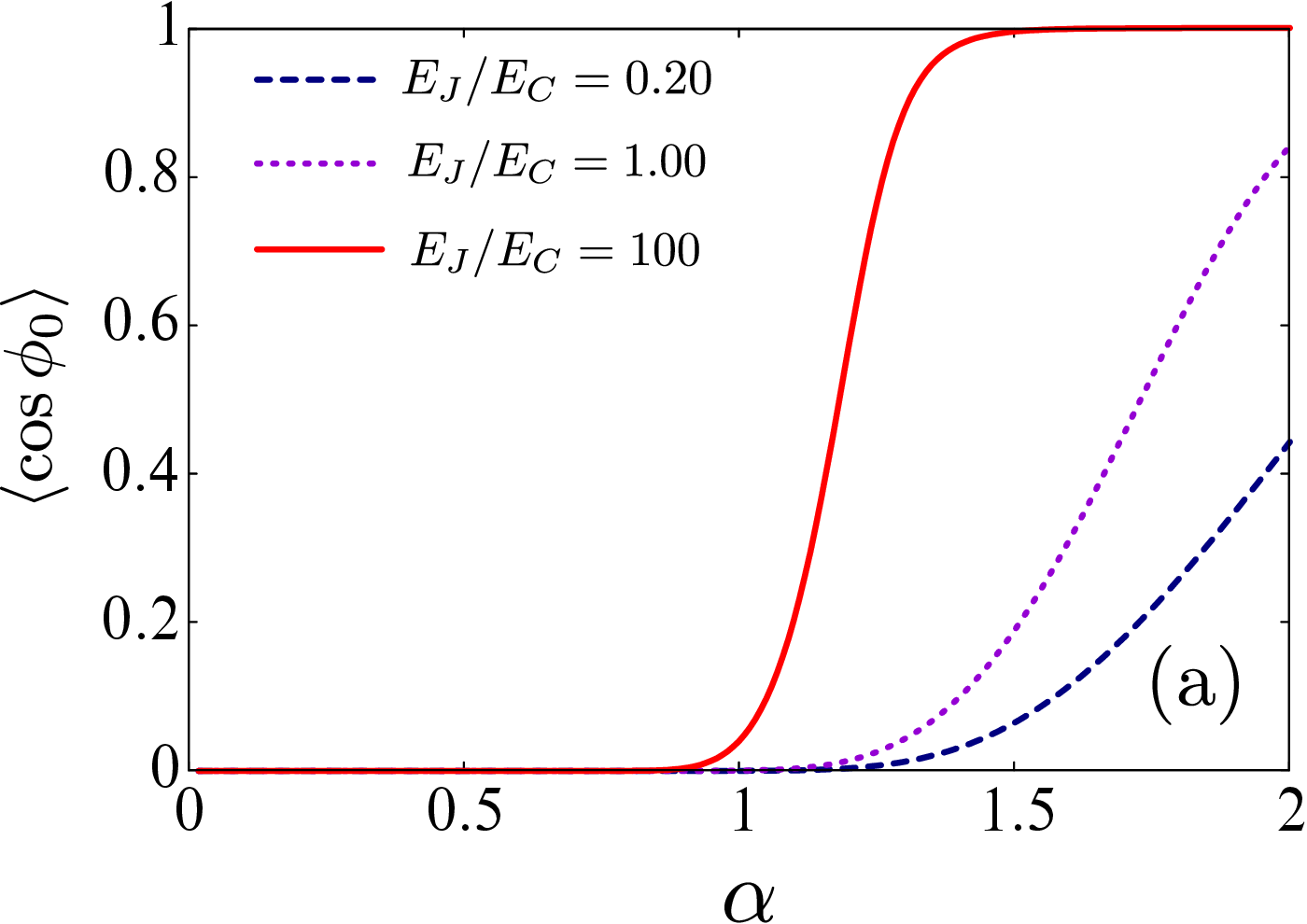}
\includegraphics[keepaspectratio,scale=0.32]{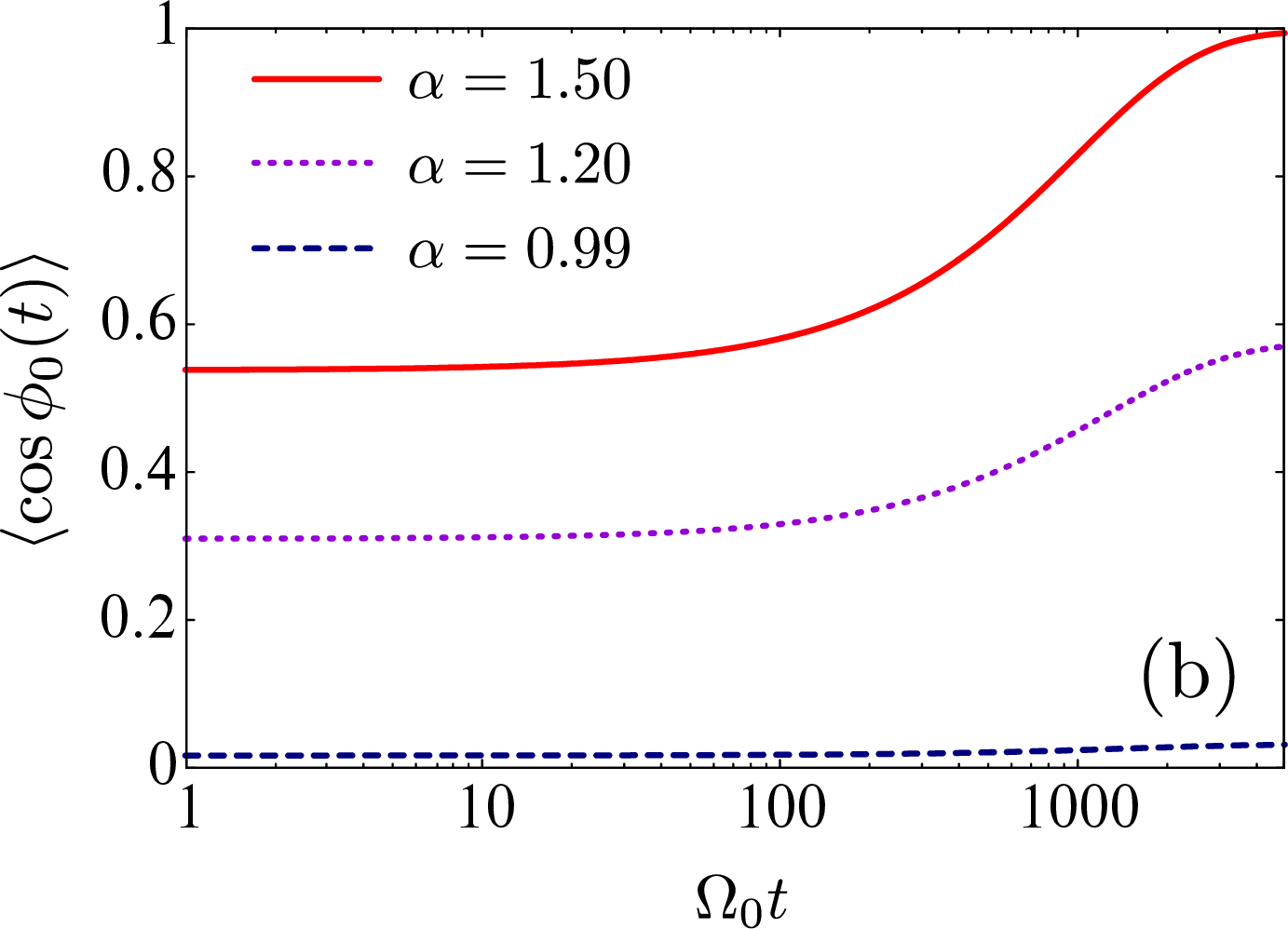}
\includegraphics[keepaspectratio,scale=0.32]{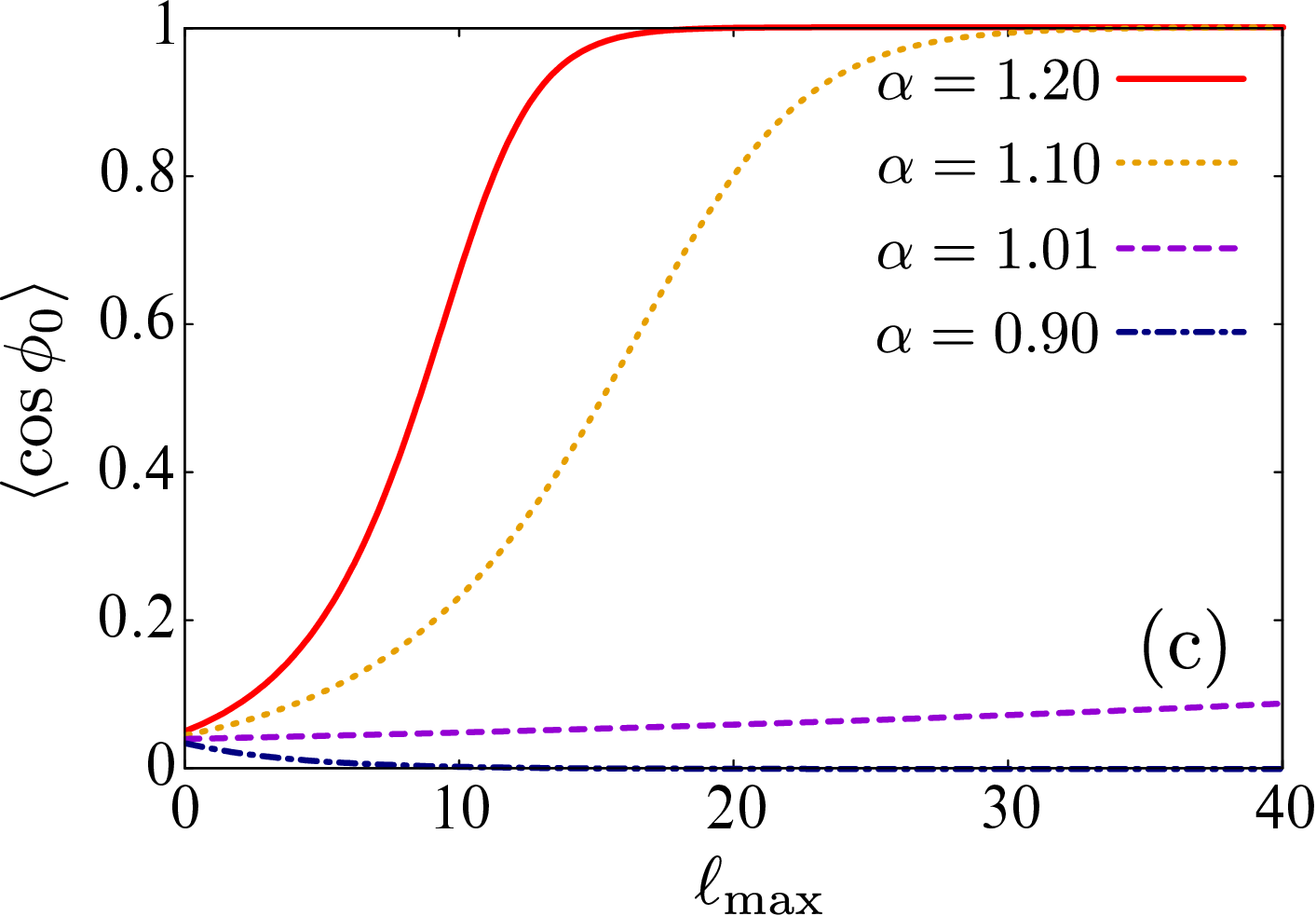}
\caption{Phase coherence computed from the Langevin equation linearizing Eq.~\eqref{langevin}. 
(a) Coherence factor in the long-time limit plotted in terms of $\alpha=2\pi\sqrt{\bar{n}\hbar^{2}/mg}$ with $g$ being the interatomic interaction strength under the initial energy ratio $E_{J}/E_{C}=0.20, 1.00, 100$ and $\ell_{\rm max}=\ln{10^{4}}$. 
(b) Time evolution of the coherence factor scaled by the bare Josephson frequency $\Omega_{0}$ with $\alpha= 1.50,1.20,0.99$ under $E_{J}/E_{C}=100$ and $\ell_{\rm max}=\ln{10^{4}}$. 
(c) RG flow of the coherence factor under $E_{J}/E_{C}=100$ with respect to the RG cutoff scale $\ell_{\rm max}$ in the long-time limit. }
\label{Figcoherence}
\end{figure}

\begin{figure}[t]
\centering
\includegraphics[keepaspectratio,scale=0.32]{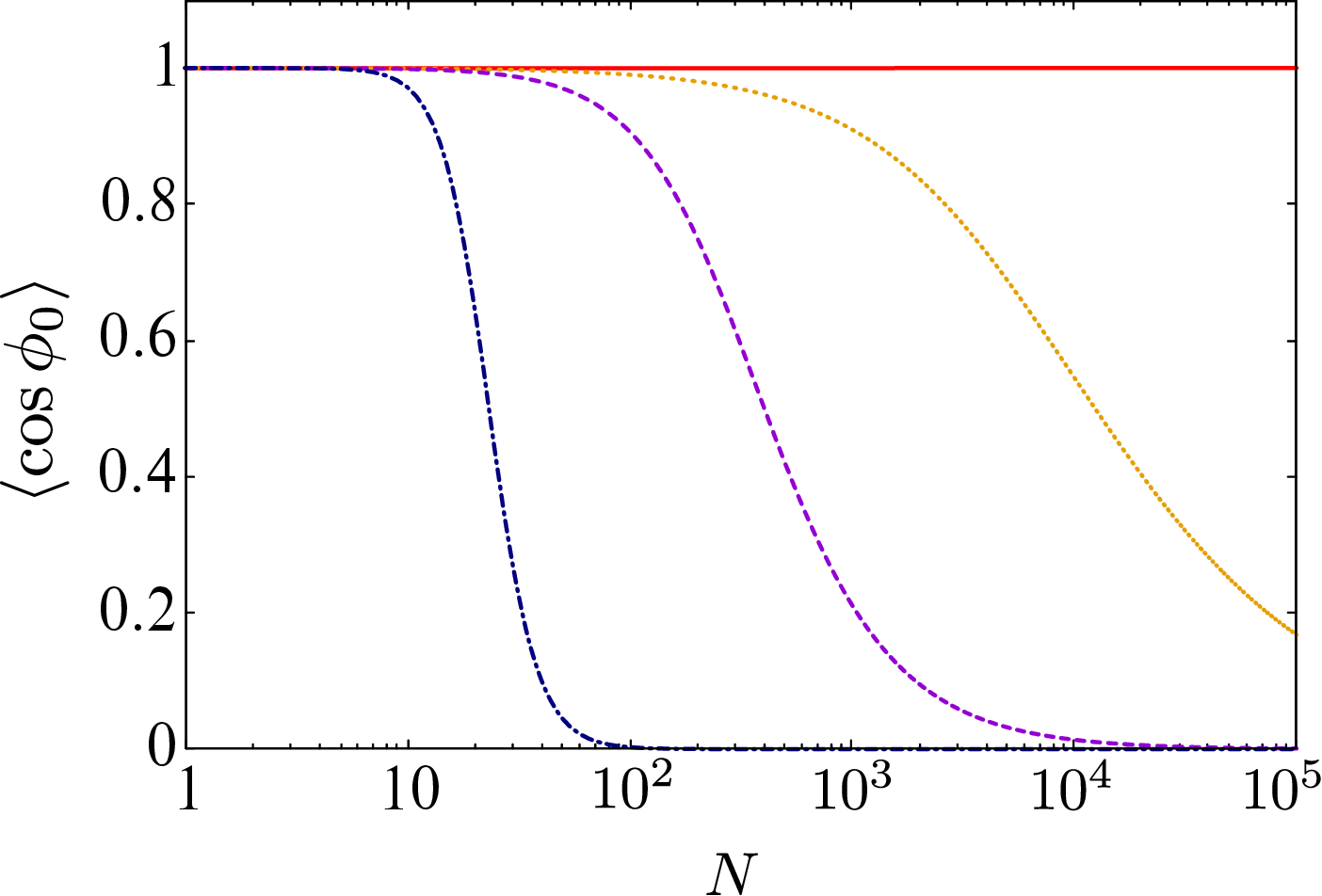}
\caption{Phase coherence in the long-time limit with $\alpha=1.1$ (solid curve), $\alpha=0.8$ (dotted curve), $\alpha=0.7$ (dashed curve), and $\alpha=0.5$ (dot-dashed curve) under $E_{J}/E_{C}=1.0$. }
\label{FigalphaNdep}
\end{figure}

The phase coherence factor $\langle\cos{\phi_{0}}\rangle$ is one of the quantities that characterize the superfluid-insulator transition \cite{fateev1997}. 
It is expected to vanish in the insulating phase while it remains nonzero in the superfluid phase. 
Reference \cite{binanti2021} derived a generalized Langevin equation for the Josephson mode as 
\beq
\bal
\dfrac{\hbar^{2}}{2E_{C}}\ddot{\phi}_{0}(t)&+\int_{0}^{t}\dd{s}\gamma[t-s;\phi_{0}(s)]\dot{\phi}_{0}(s) \\
&+\frac{E_{J}}{1+N}\sin{\phi_{0}(t)} 
=\xi(t),
\eal
\label{langevin}
\eeq
with the Gaussian noise $\xi(t)$ and $E_{J}$ renormalized under Eq.~\eqref{EJRG} and a given initial value of $E_{J}/E_{C}$. 
The damping kernel is given by $\gamma[t-s;\phi_{0}(s)]=E_{J}\sum_{n=1}^{N}\cos{[\omega_{n}(t-s)]}\cos{\phi_{0}(s)}/(1+N)$, which reduces to $h E_{J}/(\alpha E_{C})\,\delta(t-s)$ in the continuum limit at the lowest order in $\phi_{0}$ \cite{binanti2021}. 
The noise correlator at zero temperature is given by $\langle\{\xi(t),\xi(0)\}\rangle/2=\int_{-W}^{W}\dd{\omega}\chi(\omega)e^{i\omega t}$ with $\chi(\omega)=(\hbar\omega/E_{C})^{2}\hbar^{2}\abs{\omega}/2\alpha$. 
The cubic $\omega$ dependence of $\chi(\omega)$ instead of a linear one reflects the intrinsic momentum coupling. 
One can compute the coherence factor as $\langle\cos{\phi_{0}}\rangle=\cos{(\langle\phi_{0}\rangle)}\,e^{-\Delta\phi_{0}^{2}/2}$ with $\Delta\phi_{0}^{2}$ being the variance of $\phi_{0}$ by using the Gaussian property. 
Figure \ref{Figcoherence} shows the phase coherence calculated by the linearized Langevin equation. 
A small relative phase justifies the linearization of Eq.~\eqref{langevin}  and we can confirm that it also captures the features of the dissipative phase transition as we shall see in Figs.~\ref{Figcoherence}(a)-(c). 
Figure \ref{Figcoherence}(a) indicates that the coherence factor in the long-time limit grows for $\alpha>1$. 
For $\alpha\gg1$, the coherence approaches unity for any $E_{J}/E_{C}$, but the quantitative behavior depends on the initial ratio of $E_{J}/E_{C}$ and it saturates earlier with a larger ratio because the RG cutoff is chosen to be finite as $\ell_{\rm max}=\ln{10^{4}}$ based on realistic cold-atom experiments \cite{pigneur2018}. 
With an infinitely large cutoff, the curve turns into a step function that discontinuously changes to unity at $\alpha=1$ for any initial energy ratio. 
The growth dynamics of the coherence factor is displayed in Fig.~\ref{Figcoherence}(b) with three different values of $\alpha=1.50,1.20,0.99$ and the time scaled by the bare Josephson frequency $\Omega_{0}=\sqrt{2E_{J}(0)E_{C}/(1+N)\hbar^{2}}$ \cite{binanti2021}. 
It shows that the phase coherence prominently rises after a sufficiently long time around $t\sim10^{3}\Omega_{0}^{-1}$ for $\alpha>1$. 
The RG flow under $E_{J}/E_{C}=100$ in the long-time limit is plotted in Fig.~\ref{Figcoherence}(c), which clearly indicates increasing behavior for $\alpha>1$ by incrementing the RG steps. 
Conversely, the phase coherence rapidly dwindles for $\alpha<1$. 
The phase coherence represents an inductive supercurrent response carried by the ground state \cite{murani2020,joyez2011}. 
We expect that one can measure it by driving relative phase fluctuations to the junction to observe the $I$-$V$ characteristics with ultracold atoms. 
Note that the BJJ \eqref{LBJJ} is mapped to the boundary sine-Gordon model at any parameter region of $(g,J_{0},N)$. 
As a result, the Schmid-Bulgadaev phase diagram illustrated in Fig.~\ref{Figphase} is recovered and turns out to be robust against nonperturbative effects in this BJJ. 
The $(\alpha,N)$ dependence of the coherence factor in the long-time limit is plotted in Fig.~\ref{FigalphaNdep} with $E_{J}/E_{C}=1.0$ and four different values of $\alpha$. 
The total number of atoms $N$ determines the RG cutoff $\ell_{\rm max}$ and the cutoff frequency $W$. 
By increasing $N$, the phase coherence remains unity for $\alpha>1$ while it approaches zero for $\alpha<1$ with a large $N$. 
With a typical number of atoms $N \simeq 10^{4}$, the vanishing phase coherence is observed for $\alpha \lesssim 0.7$. 

Equation \eqref{alpha} implies that approaching the insulating phase requires $\Tilde{g}>4\pi^{2}$. 
Although it looks too strong coupling, this condition is feasible in a quasi-one-dimensional atomic setup. 
A path to reach the insulator phase is the renormalization of interaction by tight confinement. 
The effective interaction strength in a quasi-one-dimensional atomic system confined by tight harmonic constrictions in transverse directions is scaled as $g \propto g_{\rm 3D}/l_{\perp}^{2}$ with $g_{\rm 3D}$ being the 3D interaction strength and $l_{\perp}=\sqrt{\hbar/m\omega_{\perp}}$ being the oscillator lengths \cite{salasnich2002,nagy2015,demler2016,furutani2022}. 
Consequently, tight constrictions result in strong interaction allowing us to observe the superfluid-insulator dissipative phase transition by increasing the trap frequency $\omega_{\perp}$ as shown in Fig.~\ref{Figphase}. 
With the 3D scattering length $a_{\rm 3D}$, the critical interaction strength corresponds to $a_{\rm 3D}^{\rm c}= \pi \bar{n} l_{\perp}^{2}$, which decreases by tight constrictions $l_{\perp}\to0$. 
The superfluid state is realized in the weakly interacting regime $a_{\rm 3D}<a_{\rm 3D}^{\rm c}$ while the insulating state is favored in the strongly interacting regime $a_{\rm 3D}>a_{\rm 3D}^{\rm c}$. 

In conclusion, we proposed an alternative approach to observe the Schmid-Bulgadaev dissipative quantum phase transition with an atomic Josephson junction in a head-to-tail configuration described by the ICL model. 
It is driven by the interatomic interaction even without any synthetic dissipation. 
The crucial difference from the RSJJ is the robustness of the phase diagram against the nonperturbative effects in a head-to-tail BJJ because it can be mapped to a boundary sine-Gordon model in any parameter region in stark contrast to the RSJJ. 
Therefore, the strong suppression of the insulating phase is absent. 
We also pointed out that tight constrictions in the transverse directions responsible for the renormalization of the interaction strength allow us to reach the phase transition. 
This head-to-tail atomic junction can be realized with an atomic two-terminal system, which has been rapidly developed recently \cite{demler2016,esslinger2015,brantut2017,esslinger2019,esslinger2019spin,amico2022,esslinger2023}. 
We expect that the two-terminal setup with neutral atoms offers another platform to investigate dissipative phases in quantum systems in addition to superconducting Josephson junctions \cite{kuzmin2023,subero2023,ciuti2024}. 
An intriguing question remaining as future work is the connection between the insulating phase and the macroscopic quantum self-trapping (MQST) \cite{shenoy1999,bardin2024}, which is peculiar to atomic Josephson junctions and pins the population imbalance throughout time evolution. 
Although the charge-localized insulating phase and the MQST are similar phenomena, they are not directly connected since our analysis neglects the nonlinearity responsible for the MQST by assuming a small population imbalance, and the MQST is ruled out. 
The occurrence of the insulating phase we found has therefore distinct origin from the MQST. 
Moreover, while our analysis focused on a Josephson junction with single-component Bose gases, it is also fascinating to clarify the effects of internal degrees of freedom on the dissipative phase transition in a magnon junction \cite{nakata2014} or an atomic junction with spin degrees of freedom \cite{fromhold2013}, for instance, which would provide an insight into the interplay between dissipation and multicomponent character. 

The authors thank Y. Kawaguchi for the useful comments. 
K.F. was supported by JSPS KAKENHI (Grant No.~JP24K22858 and No.~JP24K00557) and Maki Makoto Foundation. 
L.S. is partially supported by the BIRD Project “Ultracold atoms in curved geometries” of the University of Padova; by the European Union-NextGenerationEU within the National Center for HPC, 13 Big Data and Quantum Computing [Project No.~CN00000013, CN1 Spoke 10: Quantum Computing]; by the European Quantum Flagship Project PASQuanS 2; by Iniziativa Speciﬁca Quantum of Istituto Nazionale di Fisica Nucleare; by the Project Frontiere Quantistiche within the 2023 funding program ’Dipartimenti di Eccellenza’ of the Italian Ministry for Universities and Research; by the PRIN 2022 Project Quantum Atomic Mixtures: Droplets, Topological Structures, and Vortices.


\begin{thebibliography}{99}

\bibitem{schmid1983} A. Schmid, 
Diffusion and Localization in a Dissipative Quantum System, 
Phys. Rev. Lett. {\bf 51}, 1506 (1983). 

\bibitem{bulgadaev1984} S. Bulgadaev, 
Phase diagram of a dissipative quantum system, 
JETP Lett. {\bf 39}, 264 (1984). 

\bibitem{caldeira1983} A. O. Caldeira and A. J. Leggett, 
Quantum tunneling in a dissipative system, 
Ann. Phys. (N.Y.) {\bf 149}, 374 (1983). 

\bibitem{koch1980} R. H. Koch, D. J. Van Harlingen, and J. Clarke, 
Quantum-Noise Theory for the Resistively Shunted Josephson Junction, 
Phys. Rev. Lett. {\bf 45}, 2132 (1980); 
Measurements of quantum noise in resistively shunted Josephson junctions, 
Phys. Rev. B {\bf 26}, 74 (1982).

\bibitem{kf2021} K. Furutani and L. Salasnich, 
Quantum and thermal fluctuations in the dynamics of a resistively and capacitively shunted Josephson junction, 
Phys. Rev. B {\bf 104}, 014519 (2021). 

\bibitem{murani2020} A. Murani, N. Bourlet, H. le Sueur, F. Portier, C. Altimiras, D. Esteve, H. Grabert, J. Stockburger, J. Ankerhold, and P. Joyez, 
Absence of a Dissipative Quantum Phase Transition in Josephson Junctions, 
Phys. Rev. X {\bf 10}, 021003 (2020).

\bibitem{masuki2022} K. Masuki, H. Sudo, M. Oshikawa, and Y. Ashida, 
Absence versus Presence of Dissipative Quantum Phase Transition in Josephson Junctions, 
Phys. Rev. Lett. {\bf 129}, 087001 (2022). 

\bibitem{yokota2023} T. Yokota, K. Masuki, and Y. Ashida, 
Functional-renormalization-group approach to circuit quantum electrodynamics, 
Phys. Rev. A {\bf 107}, 043709 (2023). 

\bibitem{troyer2005} P. Werner and M. Troyer, 
Efficient Simulation of Resistively Shunted Josephson Junctions, 
Phys. Rev. Lett. {\bf 95}, 060201 (2005). 

\bibitem{werner2007} S. L. Lukyanov and P. Werner, 
Resistively shunted Josephson junctions: quantum field theory predictions versus Monte Carlo results, 
J. Stat. Mech.: Theory Exp. (2007) P06002. 

\bibitem{dupuis2023} R. Daviet and N. Dupuis, 
Nature of the Schmid transition in a resistively shunted Josephson junction, 
Phys. Rev. B {\bf 108}, 184514 (2023). 

\bibitem{snyman2023} T. S{\'e}pulcre, S. Florens, and I. Snyman, 
Comment on “Absence versus Presence of Dissipative Quantum Phase Transition in Josephson Junctions”,
Phys. Rev. Lett. {\bf 131}, 199701 (2023). 

\bibitem{masuki2023rep} K. Masuki, H. Sudo, M. Oshikawa, and Y. Ashida, Reply to ‘Comment on “Absence versus presence of dissipative quantum phase transition in Josephson junctions", 
Phys. Rev. Lett. {\bf 131}, 199702 (2023). 

\bibitem{joyez2023} C. Altimiras, D. Esteve, \c{C}. Girit, H. le Sueur, and P. Joyez, 
Absence of a dissipative quantum phase transition in Josephson junctions: Theory, 
arXiv:2312.14754. 

\bibitem{glazman2024} M. Houzet, T. Yamamoto, and L. I. Glazman, 
Microwave spectroscopy of the Schmid transition, 
Phys. Rev. B {\bf 109}, 155431 (2024). 

\bibitem{minguzzi2018} J. Polo, V. Ahufinger, F. W. J. Hekking, and A. Minguzzi, 
Damping of Josephson Oscillations in Strongly Correlated One-Dimensional Atomic Gases, 
Phys. Rev. Lett. {\bf 121}, 090404 (2018).

\bibitem{binanti2021} F. Binanti, K. Furutani, and L. Salasnich, 
Dissipation and fluctuations in elongated bosonic Josephson junctions, 
Phys. Rev. A {\bf 103}, 063309 (2021). 

\bibitem{furutani2023} K. Furutani and L. Salasnich, 
Fokker-Planck equations for a trapped particle in a quantum-thermal Ohmic bath: general theory and applications to Josephson junctions, 
AAPPS Bull. {\bf 33}, 19 (2023). 

\bibitem{ashida2021} Y. Ashida, A. Imamoglu, and E. Demler, 
Cavity Quantum Electrodynamics at Arbitrary Light-Matter Coupling Strengths, 
Phys. Rev. Lett. {\bf 126}, 153603 (2021); 
Nonperturbative waveguide quantum electrodynamics, 
Phys. Rev. Research {\bf 4}, 023194 (2022). 

\bibitem{muramatsu1985} F. Guinea, V. Hakim, and A. Muramatsu, 
Diffusion and Localization of a Particle in a Periodic Potential Coupled to a Dissipative Environment, 
Phys. Rev. Lett. {\bf 54}, 263 (1985). 

\bibitem{zwerger1985} M. P. A. Fisher and W. Zwerger, 
Quantum Brownian motion in a periodic potential, 
Phys. Rev. B {\bf 32}, 6190 (1985). 

\bibitem{fisher1992} C. L. Kane and M. P. A. Fisher, 
Transmission through barriers and resonant tunneling in an interacting one-dimensional electron gas, 
Phys. Rev. B {\bf 46}, 15233 (1992). 

\bibitem{nagaosa1993} A. Furusaki and N. Nagaosa, 
Single-barrier problem and Anderson localization in a one-dimensional interacting electron system, 
Phys. Rev. B {\bf 47}, 4631 (1993). 

\bibitem{saleur1995} P. Fendley, A. W. W. Ludwig, and H. Saleur, 
Exact Conductance through Point Contacts in the $\nu=1/3$ Fractional Quantum Hall Effect, 
Phys. Rev. Lett. {\bf 74}, 3005 (1995).  

\bibitem{altlandsimons} A. Altland and B. Simons, {\it Condensed Matter Field Theory} (Cambridge University Press, Cambridge, 2010).

\bibitem{nagaosaQFT} N. Nagaosa, {\it Quantum Field Theory in Condensed Matter Physics} (Springer, Berlin, 2013). 

\bibitem{fateev1997} V. Fateev, S. Lukyanov, A. Zamolodchikov, and A. Zamolodchikov, 
Expectation values of boundary fields in the boundary sine-Gordon model, 
Phys. Lett. B {\bf 406}, 83 (1997). 

\bibitem{pigneur2018} M. Pigneur, T. Berrada, M. Bonneau, T. Schumm, E.
Demler, and J. Schmiedmayer, 
Relaxation to a Phase-Locked Equilibrium State in a One-Dimensional Bosonic Josephson Junction, 
Phys. Rev. Lett. {\bf 120}, 173601 (2018). 

\bibitem{joyez2011} I. Safi and P. Joyez, 
Time-dependent theory of nonlinear response and current fluctuations, 
Phys. Rev. B {\bf 84}, 205129 (2011). 

\bibitem{salasnich2002} L. Salasnich, A. Parola, and L. Reatto, 
Effective wave equations for the dynamics of cigar-shaped and disk-shaped Bose condensates, 
Phys. Rev. A {\bf 65}, 043614 (2002); 
Condensate bright solitons under transverse confinement, 
Phys. Rev. A {\bf 66}, 043603 (2002). 

\bibitem{nagy2015} M. Kan{\'a}sz-Nagy, E. A. Demler, and G. Zar{\'a}nd, 
Confinement-induced interlayer molecules: A route to strong interatomic interactions, 
Phys. Rev. A {\bf 91}, 032704 (2015).

\bibitem{demler2016} M. Kan{\'a}sz-Nagy, L. Glazman, T. Esslinger, and E. A. Demler, 
Anomalous Conductances in an Ultracold Quantum Wire, 
Phys. Rev. Lett. {\bf 117}, 255302 (2016).

\bibitem{furutani2022} K. Furutani and L. Salasnich, 
Superfluid properties of bright solitons in a ring, 
Phys. Rev. A {\bf 105}, 033320 (2022). 

\bibitem{esslinger2015} S. Krinner, D. Stadler, D. Husmann, J.-P. Brantut, and T. Esslinger, 
Observation of quantized conductance in neutral matter, 
Nature {\bf 517}, 64-67 (2015). 

\bibitem{brantut2017} S. Krinner, T. Esslinger, and J.-P. Brantut, 
Two-terminal transport measurements with cold atoms, 
J. Phys.: Condens. Matter {\bf 29}, 343003 (2017). 

\bibitem{esslinger2019} L. Corman, P. Fabritius, S. H{\"a}usler, J. Mohan, L. H. Dogra, D. Husmann, M. Lebrat, and T. Esslinger, 
Quantized conductance through a dissipative atomic point contact, 
Phys. Rev. A {\bf 100}, 053605 (2019). 

\bibitem{esslinger2019spin} M. Lebrat, S. H{\"a}usler, P. Fabritius, D. Husmann, L. Corman, and T. Esslinger, 
Quantized Conductance through a Spin-Selective Atomic Point Contact, 
Phys. Rev. Lett. {\bf 123}, 193605 (2019). 

\bibitem{amico2022} L. Amico, D. Anderson, M. Boshier, J.-P. Brantut, L.-C. Kwek, A. Minguzzi, and W. von Klitzing, 
Colloquium: Atomtronic circuits: From many-body physics to quantum technologies, 
Rev. Mod. Phys. {\bf 94}, 041001 (2022). 

\bibitem{esslinger2023} M.-Z. Huang, J. Mohan, A.-M. Visuri, P. Fabritius, M. Talebi, S. Wili, S. Uchino, T. Giamarchi, and T. Esslinger, 
Superfluid Signatures in a Dissipative Quantum Point Contact, 
Phys. Rev. Lett. {\bf 130}, 200404 (2023). 

\bibitem{kuzmin2023} R. Kuzmin, N. Mehta, N. Grabon, R. A. Mencia, A. Burshtein, M. Goldstein, and V. E. Manucharyan, 
Observation of the Schmid-Bulgadaev dissipative quantum phase transition, 
arXiv:2304.05806. 

\bibitem{subero2023} 
D. Subero, O. Maillet, D. S. Golubev, G. Thomas, J. T. Peltonen, B. Karimi, M. Mar{\'i}n-Su{\'a}rez, A. L. Yeyati, R. S{\'a}nchez, S. Park, and J. P. Pekola, 
Bolometric detection of Josephson inductance in a highly resistive environment, 
Nat. Commun. {\bf 14}, 7924 (2023). 

\bibitem{ciuti2024} L. Giacomelli and C. Ciuti, 
Emergent quantum phase transition of a Josephson junction coupled to a high-impedance multimode resonator, 
Nat. Commun. {\bf 15}, 5455 (2024). 

\bibitem{shenoy1999} S. Raghavan, A. Smerzi, S. Fantoni, and S. R. Shenoy, 
Coherent oscillations between two weakly coupled Bose-Einstein condensates: Josephson effects, $\pi$ oscillations, and macroscopic quantum self-trapping, 
Phys. Rev. A {\bf 59}, 620 (1999). 

\bibitem{bardin2024} A. Bardin, F. Lorenzi, and L. Salasnich, 
Quantum fluctuations in atomic Josephson junctions: the role of dimensionality, 
New J. Phys. {\bf 26}, 013021 (2024). 

\bibitem{nakata2014} K. Nakata, K. A. van Hoogdalem, P. Simon, and D. Loss, 
Josephson and persistent spin currents in Bose-Einstein condensates of magnons, 
Phys. Rev. B {\bf 90}, 144419 (2014). 

\bibitem{fromhold2013} T. W. A. Montgomery, W. Li, and T. M. Fromhold, 
Spin Josephson Vortices in Two Tunnel-Coupled Spinor Bose Gases, 
Phys. Rev. Lett. {\bf 111}, 105302 (2013). 

\end{thebibliography}
\end{document}